\documentclass[journal=jacsat,manuscript=article]{achemso}

\usepackage[version=3]{mhchem} 
\usepackage[utf8]{inputenc}
\usepackage{amsmath}
\usepackage{bm}
\usepackage{breqn}
\usepackage{multicol}
\usepackage[table,xcdraw]{xcolor}
\usepackage{color}

\newcommand{\bmu}{\boldsymbol{\mu}}
\newcommand{\balpha}{\bm{\alpha}}
\newcommand{\bA}{\mathbf{A}}
\newcommand{\bB}{\mathbf{B}}
\newcommand{\bC}{\mathbf{C}}
\newcommand{\bD}{\mathbf{D}}
\newcommand{\bE}{\mathbf{E}}
\newcommand{\bT}{\mathbf{T}}
\newcommand{\br}{\mathbf{r}}
\newcommand{\bp}{\mathbf{p}}
\newcommand{\bP}{\mathbf{P}}

\renewcommand{\d}[1]{\ensuremath{\operatorname{d}\!{#1}}}

\author{Félix Aviat}
\affiliation{Sorbonne Universités, UPMC Univ. Paris 06, UMR7616, Laboratoire de Chimie Théorique, F-75005, Paris, France}
\author{Louis Lagardère}
\email{louis.lagardere@upmc.fr}
\altaffiliation{Sorbonne Universités, UPMC Univ. Paris 06, Institut des Sciences du Calcul et des Données, F-75005, Paris, France}
\author{Jean-Philip Piquemal}
\email{jpp@lct.jussieu.fr}
\affiliation[UPMC]
{Sorbonne Universités, UPMC Univ. Paris 06, UMR7616, Laboratoire de Chimie Théorique, F-75005, Paris, France}
\alsoaffiliation{Institut Universitaire de France,Paris Cedex 05, 75231, France}
\alsoaffiliation{Department of Biomedical Engineering, The University of Texas at Austin,Austin, Texas, 78712, United States}

\title[FastGradTCG]
  {The Truncated Conjugate Gradient (TCG), a Non-iterative/Fixed-cost Strategy for Computing Polarization in Molecular Dynamics: Fast Evaluation of Analytical Forces}

\abbreviations{IR,NMR,UV}

\begin{document}


%
\begin{abstract}
 In a recent paper (J. Chem. Theory. Comput., 2017, 13, 180-190) we proposed the Truncated Conjugate Gradient (TCG) approach to compute the polarization energy and forces in polarizable molecular simulations.
 The method consists in truncating the Conjugate Gradient algorithm at a fixed predetermined order leading to a fixed computational cost and can thus be considered "non-iterative".
 This gives the possibility to derive analytical forces avoiding the usual energy conservation (\textit{i.e.} drifts) issues occurring with iterative approaches.
 A key point concerns the evaluation of the analytical gradients, which is more complex than with an usual solver.
 In this paper, after reviewing the present state of the art of polarization solvers, we detail a viable strategy for the efficient implementation of the TCG gradients calculation.
 The complete cost of the approach is then mesured as it is tested using a multi-timestep scheme and compared to timings using usual iterative approaches. 
 We show that the TCG methods is more efficient than traditional techniques, making it a method of choice for future long molecular dynamics simulations using polarizable force fields where energy conservation matters. We detail the various steps required for the implementation of the complete method by software developers.
%
%
\end{abstract}

\section{Introduction}

Polarizable force fields simulations using point dipoles models are not slow anymore. Indeed, in recent years, the computational cost of the explicit evaluation of the many-body polarization energy and associated forces has been significantly reduced using state of the art mathematical techniques.
More precisely, the bottleneck of such approaches is the mandatory resolution of a large set of linear equations (\textit{i.e.} requiring a matrix inversion) whose size depends on the number of polarizable sites, which is very large in practice (for example up to several tens of thousand of atoms for medium sized proteins in water).
Therefore, direct matrix inversion approaches are unfeasible and one has to resort to iterative methods\cite{Lipparini2014ScalableComputations} such as the Preconditioned Conjugate Gradient (PCG) or the Jacobi/Direct Inversion of the Iterative Subspace (JI/DIIS).
Both methods have the advantages to ensure convergence and to be compatible with a massively parallel implementation \citep{Lagardere2015ScalableEwald} coupled to Smooth Particle Mesh Ewald  (SPME)\cite{essmann1995smooth}, enabling the possibility to tackle large systems of interest that range from materials to biophysics.
However, iterative techniques have to address two aspects simultaneously: a low computational cost and a high accuracy on both energy \emph{and} forces. {\color {blue}But the standard way of computing the forces assumes that the dipoles are fully converged and thus these forces are not the exact opposite of the gradient of the polarization energy.}
This means that to avoid energy drifts, users have to enforce the quality of the non-analytical forces by choosing a tighter convergence criterion of $10^{–5}$ to $10^{–8}$ Debye for the dipoles, leading to a strong increase of the number of iterations required to reach convergence.
This degrades the computational efficiency of the solvers, limiting the use of molecular dynamics with polarizable force fields.
In that context, several strategies have been explored to prevent this drift while ensuring accurate results and a low computational overhead.

In this paper, we review the present status of the polarization solvers before introducing the Truncated Conjugate Gradient, a method introduced in ref. \citenum{Aviat2016TruncatedSimulations} to propose an efficient solution to these challenges. We then address the issue of the fast computation of the analytical gradients for TCG by presenting a general way to formulate the TCG polarization forces. Analytical formulas are given for the TCG1 and the TCG2 methods, as well as for their refinements with the use of a {\color {blue} preconditioner } and peek steps \cite{Aviat2016TruncatedSimulations}. {\color {blue}Indeed as a preconditioner improves the convergence of the polarization computation,  a peek step allows to perform a additional but inexpensive Jacobi/Picard pseudo-iteration that does not requires any matrix-vector product as it uses the available residual obtained from the TCG process.} Finally, timings to compute these forces in a production context of a Respa integrator are given and compared to the ones obtained with standard iterative solvers and different level of convergence as well as different predictor guesses for these solvers.

\section{Polarization solvers: present status}

Several iterative solvers applied to the polarization equations have been presented and tested, such as the Jacobi Over Relaxation method (JOR), the (Preconditioned) Conjugate Gradient, the Jacobi/DIIS method (see references \citenum{Lipparini2014ScalableComputations} and   \citenum{Lagardere2015ScalableEwald}) or the recently introduced potentially faster Divide and Conquer block-Jacobi/DIIS method \cite{Nocito2017FastFields}.

Considering an iterative solver, several techniques can be used to reduce the computational cost to reach convergence by reducing the number of necessary iteration to do so. In the context of Krylov methods such as the Conjugate Gradient, it is for example possible to use a preconditioner.
It consists in choosing a matrix $\bP$ such that $\bP^{-1}$ is close to $\bT^{-1}$ (where $\bT$ is the polarization matrix to be inverted, presented in the third section of the paper) and in applying the iterative method to the modified linear system where the matrix and the right hand side are multiplied by $\bP^{-1}$.
The convergence of the solver is then accelerated because of the clustering of the eigenvalues of the matrix $\bP^{-1}\bT$ \cite{Lipparini2014ScalableComputations} .
Efficient preconditioners for the polarization equations have been designed, such as the ones proposed by Wang and Skeel \cite{Skeel_JCP_PPD} which provide a reduction in the number of iteration to reach convergence up to 10 to 20 percent, depending on the system (\textit{i.e.} on the condition number of the matrix that one needs to invert).

Another way to improve convergence of an iterative solver is to chose an initial "predictor" guess as close as possible to the actual solution of the linear equations.
This guess can be constructed using information from one or a few of the past values of the dipoles.
The most naive way to do so is to chose the value of the dipoles at the previous timestep (previous guess) but more elaborate and efficient strategies have been designed such as Kolafa's Always Stable Predictor Corrector (ASPC) \cite{kolafa2004time} or Skeel's Least Square Predictor Corrector (LSPC) \cite{Skeel_JCP_PPD}, that can reduce the number of iterations required to reach convergence up to a factor two in a standard production context \cite{Lipparini2014ScalableComputations} .
Nevertheless, these two ways to construct initial guesses lose their efficiency when one uses larger timesteps, as it the case with the RESPA (Reversible reference System 
Propagator Algorithm) multiple timestep integrator \cite{tuckerman1992reversible} (instabilities occur when such predictors are used with time steps larger than 2~fs).

Note that the two refinements (preconditioning and choosing the initial guess of the solver wisely) can be coupled without problem.

In the same spirit, it is also possible to speed up convergence by introducing an extended Lagrangian scheme to propagate a set of dipoles that are used as initial guess to standard iterative solvers (iEL/SCF or Extended Lagrangian Self-Consistent Field, see ref. \citenum{albaugh2015efficient}).
This approach, derived from ab initio MD, significantly reduces the number of iterations of the solver (by the same order of magnitude as the ASPC predictor) but requires to use an additional thermostat in order to prevent energy flows between the degrees of freedom.

However, whatever the different speedups strategies applied of the popular iterative production methods such as PCG or JI/DIIS, they still suffer from an important drawback in link to the way the associated forces are computed.
Indeed, they do not address the polarization energy drifting issues that will be encountered in long simulations of large non-homogeneous complexes, such as proteins in water or highly charged ionic liquids. In such case, the mathematical problem, \textit{i.e.} the matrix inversion, is costlier to solve as the polarization matrix itself is worse conditioned than in simple bulk water.
Therefore, to ensure stability of very long timescale simulations towards microseconds where errors accumulate, they should all employ a tighter dipole convergence criterion ($10^{-7}$ to $10^{-8}$~D) leading to a higher number of iterations than usually discussed in benchmarks  for short simulations, where the $10^{-5}$~D standard is employed, effectively causing really degraded real life performances.

Another set of methods address this issue by considering analytical formulas for the polarization energy.

The first idea in that direction was introduced by Wang and Skeel \cite{Wang2005FastSimulations}, who used Chebyshev polynomials to get analytical expressions of the polarization energy and its derivatives, which automatically ensures that the source of the energy drift previously evoked is removed.
Unfortunately, the approach provided energy surfaces that were too far from the ones obtained with tightly converged iterative method and was thus not further investigated. 
Significant progresses were recently made in the same direction by Simmonett et al. \cite{simmonett2015efficient} who proposed a revisitation of Wang's proposal through the ExPT (Extrapolated Perturbation Theory) perturbation approach, which is equivalent to the truncation of the Jacobi iterative method at a predetermined order combined with the use of a few parameters.

If the parametric aspect of their approach initially limited its global applicability to any type of systems, the authors recently improved their method which is now denoted OPT3 (OPT=Orders of Perturbation Theory)\cite{simmonett2016empirical} by pushing it to higher order of perturbation and providing a systematic way for the parametrization, extending the applicability of the method.
One advantage of the approach is its reduced cost compared to the best iterative approaches.

Alternatively, one can also consider the actual induced dipoles as new degrees of freedom and build an extended Lagrangian defining the way to propagate them during the dynamics without any SCF cycles \cite{albaugh2017accurate} .
The first results using this strategy are promising and the method indeed does not require any iteration. On the performance side, one could argue that using a production PCG solver with a $10^{-5}$~D convergence threshold, a RESPA integrator with a 2~fs time step for the non bonded forces coupled to Kolafa's ASPC is twice faster than the sequential iEL/0-SCF method with a 1~fs time step \cite{albaugh2017accurate} .
Nevertheless, this PCG speed advantage is only "apparent" as it does not solve the energy drift issue for long time scales whereas the iEl/0-SCF method has been shown to have improved energy conservation properties. This nice improvement is due to the use of thermostats and therefore, iEL/0-SC unfortunately suffers from the drawbacks of any extended Lagrangian approach that can not use time steps larger than 1~fs. \cite{Skeel_JCP_PPD} As we stated before, if iterative methods do not have any theoretical upper limit to the time step they can be used with\cite{Skeel_JCP_PPD}, it requires not to use information from the past such as predictor-correctors, removing such speed advantage when using RESPA.  

As we see from this discussion, the question of which method to adopt is complex as it appears difficult to combine all possible improvements.
 
In fact, we can state that reducing the computational cost of an iterative method to compute the polarization energy and forces always come with degraded energy conservation.
Energy conservation is tricky as it depends on the chemical nature of the system (charged or not, homogeneous or not).
For example, polarization of bulk water systems requires less iterations to converge with PCG solvers. On the other hand, the ExPT method behaves poorly for the ionic liquid system that will be studied in section 4 \cite{Aviat2016TruncatedSimulations} and the Jacobi method does not even converge in that case . 

A major difficulty to compute the polarization energy and its gradient for future microseconds simulations is to offer a non-empirical strategy applicable to any kind of systems, embodying the following properties.

Indeed, such a method should be systematically improvable in order to allow the user to set the accuracy of the simulation depending on its goal. For example, the simple Jacobi method has been shown not to converge in several cases \cite{Lagardere2015ScalableEwald} and adding iterations would not improve the results.
It should show good conservation of the total energy during a microcanonical simulation, ensuring good accuracy on the forces driving the dynamics. 
It should also be non parametric to provide a close reproduction of any type of potential energy surfaces, without having to resort to force-field models reparametrization.
In practice, a polarization scheme should also be affordable with a computational cost as reduced as possible. It should allow to use larger time steps through multiple timestep schemes such as RESPA.  In the end, the selected criterion to compare computational efficiencies of the various schemes should be the global cost of computing both energy and derivatives with similar energy conservation capabilities for a given trajectory length.

\section{TCG : context}

To address all these required features we recently introduced a non-empirical and non-iterative strategy denoted the Truncated Conjugate Gradient (TCG)\cite{Aviat2016TruncatedSimulations}.
{\color {blue} TCG is derived by explicitly writing down all numerical operations of a finite number of Conjugate Gradient cycles of iteration which can be user-chosen (be TCG-n, n=1,3).}
As the number of operations in the TCG approach is fixed once and for all, it is possible to derive an exact analytical expression of the gradient of the energy like in ExPT/OPT3 \cite{simmonett2016empirical}, avoiding by construction any energy drift in microcanonical simulations and thus ensuring energy conservation in that context.
The higher the TCG level is, the higher its accuracy is, as TCG inherits from the properties of the Conjugate Gradient and {\color {blue} benefits} from the fact that it is a Krylov method in which the associated error is monotonically reduced at each iteration.
It can be shown in that context that the CG-method is mathematically optimal, meaning that it minimizes exactly the polarization energy on the so-called Krylov subspaces at each iteration and therefore guarantees that the number of the required matrix-vector products (1 per iteration in any iterative approach) are reduced to a minimum compared to other iterative methods.
Moreover, the TCG accuracy can be improved at negligible costs (\textit{i.e. }without any additional matrix-vector product): (i) by using preconditioners as presented above leading to the Truncated Preconditioned Conjugate Gradient (TPCG); (ii) by using the residue of the final CG step, available without any additional cost, to perform an additional “peek” iteration, equivalent to one step of Jacobi Over Relaxation (JOR) with a relaxation parameter which can be found adaptively. 

Overall, the TCG approach was found to accurately reproduce energy surfaces at a reduced computational cost providing analytical forces. 
As it does not rely on history, it does not suffer from MD perturbations such as the ones arising when predictor guesses, {\color {blue} which break } the time-reversibility of the simulation, are used in polarization solvers. It is for the same reasons compatible with the use of large timestep with multi-timesteps integrators.
Also, being based on the Conjugate Gradient and thus relying essentially on matrix vector products and computation of electric fields, it can replace standard solvers in a regular implementation including linear scaling ones using Smooth Particle Mesh Ewald. Furthermore, it does not require additional advanced thermostating nor any additional parameter. 

The purpose of this paper is to address one delicate point which is the main bottleneck of the TCG method: the complex derivation of its gradients.
If TCG answers all the desired discussed properties for a polarization solver, a naive derivation of the energy gradients can lead to an undesired additional computational cost, while the method  should remain analytical, accurate but cheap as well. 
The goal here is to detail a strategy enabling a fast computation of the analytical gradients that would allow developers to efficiently implement the TCG approach in the software of their choice.
We will first present the technical aspect of TCG and its notations, then we will detail the optimal computation of gradients in a form that could be implemented by developers.

\section{TCG : notations}

We will place ourselves in the context of the AMOEBA force field \cite{Ponder2007CurrentField} and consider a system of $N$ atoms, each embodying a multipole expansion (up to quadrupoles) as permanent charge density and a polarizability tensor $\alpha_i$. 
 We will {\color {blue} denote $\bE$ } as the $3N$ vector gathering all electric fields $\vec{E_i}$ created by the permanent charge density at atomic position $i$, and $\bmu$ is the equivalent $3N$ vector gathering the induced dipoles experienced at each atomic site.
 $\bT$ is the $3N\times 3N$ polarization matrix, defined by block as follows. It bears the $3\times 3$ polarizability tensors $\alpha_i$ along its diagonal block, and the interaction between the $i$th and $j$th dipole is represented as the $T_{ij}$ tensor. 
     \[ \bT = \left(
     \begin{array}{ccccc}
      \alpha_1^{-1} & -T_{12} & -T_{13} & \ldots & -T_{1N}      \\
      -T_{21} & \alpha_2^{-1} & -T_{23} & \ldots & -T_{2N}      \\
      -T_{31} & -T_{32}       & \ddots  &        &              \\
      \vdots  & \vdots        &         &        &\vdots        \\
      -T_{N1} & -T_{N2}       &         & \ldots & \alpha_N^{-1}\\
     \end{array}
     \right)
     \]
     
{\color {blue} This matrix is symmetric and positive definite. Thanks to the Thole damping of the electric field at short range, any polarization catastrophe is prevented. Indeed, the Thole damping acts on the eigenvalues. Without Thole, negative eigenvalues could be found which is a problem for Conjugate Gradient methods. \cite{Lipparini2014ScalableComputations} }

    Using these notations, the total polarization energy can be expressed as follows : 
    
  \begin{equation}
      E_{\text{pol}} = \frac12 \bmu^T \bT \bmu - \bmu^T \bE \label{eq:Epol}
  \end{equation}
    where $\bmu^T \bE$ represents the scalar product of vectors $\bmu$ and $\bE$ (also noted $\langle \bmu, \bE \rangle$).
	One can easily see that the dipole vector $\bmu$ minimizing (\ref{eq:Epol}) verifies the following linear system:
    \begin{equation}
    	\bT \bmu = \bE
    \end{equation}
	giving the minimized polarization energy: 
	\begin{equation}
    	E_{\text{pol}}  = - \frac12 \bmu^T \bE 
    \end{equation}
    
    As explained earlier, the TCG method that we use to solve this equation, derives from the Conjugate Gradient algorithm. 
    It uses three vectors upon starting : the guess $\bmu_0$, the initial residual $\br_0 = \bT \bmu_0 - \bE $, and an initial descent direction $\bp_0$ that we set to be equal to $\br_0$.
    It  reads as follows: 
   \begin{eqnarray}
       \left\{
          \begin{array}{lllll}
             \gamma_i = \frac{\br_i^\bT \br_i}{\bp_i^T \bT \bp_i} \\
             \bmu_{i+1} = \bmu_i + \gamma_i \bp_i \\
             \br_{i+1} = \br_i - \gamma_i \bT \bp_i \\
             \beta_{i+1} = \frac{\br_{i+1}^T\br_{i+1}}{\br_i^T\br_i} \\
             \bp_{i+1} = \br_{i+1} + \beta_{i+1}\bp_i
          \end{array}
       \right.
    \end{eqnarray}
    Instead of using a convergence criterion as a condition to stop iterating, as {\color {blue} this is usually done,} one can choose to arbitrarily fix the number of iterations and to unfold a finite number of computational operations that makes it fixed cost and non-iterative, as explained above.
    This defines our Truncated Conjugate Gradient (TCG) method. 
    Besides the obvious advantage of drastically reducing the computational cost of each induced polarization calculation, it allows one to simulate perfectly stable molecular dynamics, without drift over time, as explained in Ref. \citenum{Aviat2016TruncatedSimulations}. This advantage is not limited to MD and could be exploited in Monte-Carlo simulations.
   
   The exact, total derivative of the energy with respect to the nuclear position should be:
   \begin{align}
   	   \frac{\d E_{\text{pol}}}{\d r_i} = \frac{\partial E_{\text{pol}}}{\partial \bmu} \frac{\partial \bmu}{\partial r_i} + \frac{\partial E_{\text{pol}}}{\partial r_i}  \label{eq:HFEpol}
   \end{align}
   
   When using an iterative method, the provided solution $\bmu$ is inexact (approached only), thus the energy is not perfectly minimized with respect to the dipoles (the term $\partial E_{\text{pol}} / \partial \bmu$ is not zero). One usually still makes this erroneous assumption, giving $\d E_{\text{pol}} / \d r_i = \partial E_{\text{pol}} / \partial r_i$. This leads to computing forces that do not perfectly correspond to the system, and thus to an unavoidable drift in the subsequent simulations.

    If one fixes the number of iterations, it is however possible to "unroll" the analytical formula for the final polarization vector, expressed as a function of the starting quantities ($\bmu_0$, $\br_0$). 
    Noting $\bmu_{\text{TCG}n}$ this vector, with $n$ the truncation order (\textit{i.e} the number of iterations of the algorithm), one obtains 
 the $\text{TCG}n$ family of methods that reads up to order three:  \begin{align}
       \bmu_{TCG1} = \bmu_0 + t_4 \br_0 \label{eq:mu1} \\
	   \bmu_{TCG2} = \bmu_0 + (  \gamma_1 t_2 + t_4 )\br_0 - \gamma_1 t_4 \bP_1  \label{eq:mu2} \\
	   \bmu_{TCG3}= \bmu_0 + ( t_4 + \gamma_1 t_2 + \gamma_2 + \gamma_2 \beta_2 t_2) \br_0 - ( \gamma_1 t_4 + \gamma_2 t_4 + \gamma_2 \beta_2 t_4 ) \bP_1 - \gamma_1 \gamma_2 \bP_2   \label{eq:mu3}
	\end{align}
   
All quantities used in the previous equations are defined in the {\color {blue} Appendix.}
In practice, we showed that one could stop as the TCG2 level, as it is accurate enough.

\section{Fast computation of the gradients}

	In this section, we first explain that computing the gradients of the energy, {\color {blue} even though } an analytical expression is at our disposal, is not straightforward. We then show how to pass the different hurdles encountered.
    
    Having the analytical, exact expression of the dipoles allows one to differentiate them in an equally exact manner. A formal differentiation, with a prime "$\: '\:$" denoting it, would give for the first two  orders: 
    \begin{align}
       \bmu_{TCG1}' = \bmu_0' + t_4\br_0' + t_4'\br_0  \label{eq:muTCG1'} \\
       \bmu_{TCG2}' = \bmu_0' + (t_4 + \gamma_1 t_2) \br_0' + (t_4' + \gamma_1' t_2 + \gamma_1 t_2') \br_0 + \gamma_1' t_4 \bP_1 + \gamma_1 t_4' \bP_1 + \gamma_1 t_4 \bP_1' \label{eq:muTCG2'}
    \end{align}    
    
     However, the differentiation of a $3N$ vector with respect to $3N$ spatial coordinates would build a $3N\times 3N$ matrix. This leads to three obstacles that slow down the gradient computation :
     \begin{itemize}
     	\item firstly, a scalar product of one such derivative $\bA'$ with another vector $\bB$ would lead to a $(3N)^2$ operation, which is a non-negligible cost, repeated for all products of this $<\bA' ,\bB>$ form. 
        \item Secondly, these products, when using the analytical expressions (equations \ref{eq:muTCG1'} and \ref{eq:muTCG2'}) "as is", are repeated an unnecessary number of {\color {blue} times, } effectively making this slow-down a pure stop. 
        \item Thirdly, one can see that there are two types of vectors building  $\bmu_{\text{TCG}n}$: the electric field $\bE$, but also the product of the residue with successive powers of the polarization matrix ($\br_0$, $\bT \br_0 = \bP_1$, more generally $\bT^m \br_0$, with $m$ an integer). Differentiating $\bT^m \br_0$ exhibits, amongst others, a $\bT^p \bT' \bT^q \br_0$ term (with $p$ and $q$ two integers verifying $p+q+1 = m$); computing such a $\bT .\bT' \bA $ product is equivalent to a matrix-matrix product, which is also computationally too expensive.
     \end{itemize}
     This makes a naive implementation of our method effectively unusable.
     Yet to run a classical simulation, one needs the \emph{forces}, \textit{i.e.} the gradients of the polarization energy, rather than the derivatives of the dipoles themselves. 
     What one really needs is thus the derivative of the following scalar product :
     \begin{equation} 
         E_{\text{pol}} = \frac12 < \bE, \bmu_{\text{TCG}n} > 
     \end{equation}

that is, formally, 
    \begin{equation}
        E_{\text{pol}}' = \frac{1}{2} < \bE' , \bmu_{\text{TCG}n} > + \frac12 < \bE , \bmu_{\text{TCG}n}'>  \label{eq:epolprime}
    \end{equation}

    Firstly, developing eq. (\ref{eq:epolprime}) shows all scalar products involved involve a differentiated quantity : either a differentiated matrix (like $<\bA, \bT'\bB>$), or the derivative of the field itself ($\bE'$). 
    An analogy, or dimensional analysis, allows us to compare these terms to forces, with $< \bA, \bE'>$ corresponding to a force produced by the interaction of the dipoles $\bA$ with the electric field, and $<\bB, \bT' \bC >$ to a force arising from the interaction between two sets of dipoles $\bB$ and $\bC$. 
    The expensive part of computing such quantities lies in the calculation of distances. 
    All of these forces can be computed in a single double loop (whose cost is a $O(N^2)$ for direct calculations, and $O(N \log N)$ when using SPME) to minimize the computational cost and compute the said distances only once. 
    This {\color {blue} adresses } the first hurdle evoked earlier.

    We can also reorganize the gradient computation in order to minimize the number of the expensive scalar products involving a vector and a differentiated vector, by grouping all these scalar products and performing them all at once (given three vectors $\bA$, $\bB$ and $\bC$, if one needs to compute $<\bA, \bB'> + <\bC, \bB'>$, it is much more efficient to first prepare a vector $\bD = \bA + \bC$ and then to compute $<\bD, \bB'>$). 
     This optimization, though quite simple in principle, actually requires quite involved expressions (see Annex). It is a simple solution to the second obstacle we listed.

    Thirdly, since $\bT$ is a symmetric matrix, we have $<\bT \bA, \bB> = < \bA, \bT \bB>$ for any two vectors $\bA$ and $\bB$. 	In particular, for our generic vectors $\bT^m \br_0$, 
    \begin{eqnarray}
    	<\bT^p \bT' \bT^q \br_0, \bA > = < \bT' \bT^q \br_0, \bT^p \bA>
    \end{eqnarray} 
    Considering scalar products thus allows us to get rid of the matrix-matrix ($\bT. \bT'$) products, our third hurdle.
    
    Overall, the solution to overcome our obstacles came from considering the polarization energy instead of the induced dipole themselves.
	
    To illustrate our solution, one can write the analytical formulas as follows, for the TCG at order one and two respectively :  
     \begin{dmath}
        E_{\text{pol, TCG1}}' = \frac{1}{2} \left( \langle  \br_0', a_{1,0}^{(1)} \bE + a_{1,1}^{(1)} \br_0   + a_{1,2}^{(1)} \bT \br_0     \rangle + \langle  \bT' \br_0,  a_{2,1}^{(1)} \br_0 \rangle   \right)  \label{eq:E1generic}
    \end{dmath}
     \begin{dmath}
        E_{\text{pol, TCG2}}' =\frac{1}{2} \left( \langle \bE', \bmu_{\text{TCG2}} \rangle + \langle \bmu_0' , \bE \rangle  + \langle \br_0', a_{1,0}^{(2)} \bE  + a_{1,-1}^{(2)} \bT \bE + a_{1,1}^{(2)} \br_0   + a_{1,2}^{(2)} \bT \br_0   + a_{1,3}^{(2)} \bT^2 \br_0    + a_{1,4}^{(2)} \bT^3 \br_0  \rangle  + \langle \bT' \br_0, a_{2,0}^{(2)} \bE  + a_{2,1}^{(2)} \br_0  + a_{2,2}^{(2)} \bT \br_0   + a_{2,3}^{(2)} \bT^2 \br_0  \rangle + \langle \bT' \bT \br_0, a_{3,1}^{(2)} \br_0 + a_{3,2}^{(2)} \bT \br_0   \rangle + \langle \bT' \bT^2 \br_0, a_{4,1}^{(2)} \br_0  \rangle  \right) \label{eq:E2generic}
    \end{dmath}
    where the coefficients $a_{i,j}^{(k)}$ are the result of the cumbersome derivation evoked earlier ; their explicit expression can be found in the Annex.
   
    As stated earlier in this paper, the so-called peek-step is a supplementary JOR iteration based on the last obtained residual $\br_n$. It simply improves the solution to reach the following expression :
    \begin{eqnarray}
    	\bmu_{\text{TCG}n}^{(peek)} = \bmu_{\text{TCG}n} + \omega \balpha \br_n
    \end{eqnarray}
    $\alpha$ is the relaxation parameter mentioned earlier; more precisions on its choice can be found in ref. \citenum{Aviat2016TruncatedSimulations} .
    Defining $\bmu_{\textrm{peek, TCG}n} = \omega \balpha \br_n$, the supplementary contribution of the peek step can be also written as follows :
    
      \begin{dmath}
    E_{\text{peek, TCG1}}' = \langle \bmu_{\text{peek, TCG1}}, \bE' \rangle + \langle \br_0', a_{1,\alpha0}^{(1,p)} \balpha \bE  + a_{1,1\alpha}^{(1,p)} \bT \balpha \bE  + a_{1,1}^{(1,p)} \br_0  + a_{1,2}^{(1,p)} \bT \br_0  \rangle + \langle \bT' \br_0 , a_{2,1}^{(1,p)} \br_0  + a_{2,\alpha0}^{(1,p)} \balpha \bE  \rangle
    \end{dmath}
    
    \begin{dmath}
    E_{\text{peek, TCG2}}' = \langle \bmu_{\text{peek, TCG2}}, \bE' \rangle  \\
    + \langle \br_0' , a_{1,0\alpha}^{(2,p)} \balpha \bE +  a_{1,1\alpha}^{(2,p)} \bT \balpha \bE  + a_{1,2\alpha}^{(2,p)}  \bT^2 \balpha \bE  + a_{1,1}^{(2,p)} \br_0  + a_{1,2}^{(2,p)} \bT \br_0  + a_{1,3}^{(2,p)} \bT^2 \br_0  + a_{1,4}^{(2,p)} \bT^3 \br_0  \rangle \\ 
    + \langle \bT' \br_0 ,a_{2,\alpha0}^{(2,p)} \balpha \bE  + a_{2, 1\alpha}^{(2,p)} \bT \balpha \bE  + a_{2,1}^{(2,p)} \br_0  + a_{2,2}^{(2,p)} \bT \br_0  + a_{2,3}^{(2,p)} \bT^2 \br_0  \rangle + \langle \bT' \bT \br_0 , a_{3,\alpha0}^{(2,p)} \balpha \bE  + a_{3,1}^{(2,p)} \br_0  + a_{3,2}^{(2,p)} \bT \br_0   \rangle + \langle \bT' \bT^2 \br_0, a_{4,1}^{(2,p)} \br_0  \rangle
    \end{dmath}  
    
    (the coefficients $a_{i,j}^{(k,peek)}$, as well as an explicit formula for the $\bmu_{\text{peek}}$ vectors, are reproduced in the Annex). One should then simply sum the corresponding terms to obtain the final expression for the polarization energy gradients in a computationally feasible way : for example, the scalar product $\langle \br_0', \br_0 \rangle$ {\color {blue} should now be multiplicated} by coefficient $a_{1,1}^{(1)} + a_{1,1}^{(1,p)}$ to get the correct gradients for TCG1.
    
    All these formulas have been tested and validated against gradients obtained via finite differences. 
    Such details could be useful to allow anyone to implement the fast evaluation of the forces necessary to the use of TCG. The source code of this method will be freely available in Tinker-HP version 1.1 \cite{webtinkerhp}.

	To sum up, the implementation of the gradients calculation that we propose here follows these  three steps : firstly, we compute the successive matrix-vector products to build the successive $\bT^m \br_0$ vectors needed; secondly we perform the various scalar products appearing in our analytical formulas, allowing us to assemble (through weighted sums) a second set of vectors; finally, we perform simultaneously on all these assembled vectors a "force-like" calculation. The choice to use -- or not -- a peek step only changes the assembled vectors on step two, through an extra set of coefficients as presented above.

\section{Numerical results}

In this section, we report the timings of the implementation presented above for different systems as it has been added to the software Tinker-HP .
More precisely, we report the cost of the calculation of the polarization energy and the associated forces with different methods: a standard diagonally preconditioned conjugate gradient (PCG) with a $10^{-5}$D convergence threshold, the same method with a tighter $10^{-8}$D convergence threshold (that ensures energy conservation as explained above) and the TPCG1 and the TPCG2 methods with the "direct field" \cite{Lipparini2014ScalableComputations} $\balpha \bE$  as guess $\bmu_0$ with a Jacobi peek step ($\omega$=1).
For the two PCG solver settings the average number of iterations is also reported in parenthesis.
Note that the computational cost of these two methods would be the same with any other kind of peek steps whose cost is negligible, as described in ref.\citenum{Aviat2016TruncatedSimulations} .
 For the PCG solvers, we report timings using the simple "direct field" as a guess (noted "PCG ($10^{-x}D$)" in the table) and also timings using the ASPC predictor (noted "PCG ($10^{-x}D$, ASPC)" )\cite{kolafa2004time}. These methods are timed in the nowadays standard context of the RESPA integrator \cite{tuckerman1992reversible} used with a 2 fs time step for the non bonded forces.

The systems that are tested here are the same than in our previous work \cite{Aviat2016TruncatedSimulations}: three solvated protein droplets (the HIV nucleocapsid ncp7 made of 18518 atoms, the ubiquitin made of 9737 atoms and the dihydrofolate reductase dhfr with 23558 atoms) and an ionic liquid, the dimethyl-imidazolium [dmim+][Cl-] (3672 atoms).
 {\color {blue} No boundary conditions are used in these tests, therefore, each matrix-vector product and force computation involved in the PCG solvers and in the TCG formulas has a $O(N^2)$ computational cost. However, these matrix-vector products can be easily re-expressed following the possible choices for the boundary conditions that will give rise to slightly different forms of the polarization matrix. For example, TCG being really close to PCG, it can either be applied in the context of the Particle Mesh Ewald \cite{Lagardere2015ScalableEwald}  \cite{darden1993particle} method with a $O(NlnN)$ cost, or using the Fast Multipole summation technique\cite{FMM} with a $O(N)$ cost.} These operations are by far the costliest in the computation of the dipoles and of the polarization forces. 
This is why we report the timings as their proportional cost compared to the PCG solver with a convergence threshold of $10^{-5}$~D and the direct field as a guess, as these proportions would be the same when using other boundary conditions. We chose these settings to be our reference.

All these (sequential) timings were obtained on an HP 620 Workstation made of Intel Xeon E5-2665 CPUs at 2.4~Ghz and were averaged over 100~ps of NVT trajectories at 300~K for the protein droplets and at 425~K for the ionic liquid.

\begin{table}[h]
\centering
\begin{tabular}{|c|c|c|c|c|}
  \hline
   & ubiquitin & ncp7 & dhfr & [dmim+][Cl-]\\
  \hline
  \rowcolor[HTML]{EFEFEF} 
  PCG ($10^{-5}$D) &100\% (8) & 100\% (8)& 100\% (8)&100\% (8)\\
  PCG ($10^{-5}$D, ASPC)&88\% (6)& 85\% (6)& 88\% (6)&84\% (5)\\
  \rowcolor[HTML]{EFEFEF} 
  PCG ($10^{-8}$D) & 136\% (15) & 138\% (15) & 143\% (16) & 138\% (15)\\
  PCG ($10^{-8}$D, ASPC)&125\% (13) & 127\% (13) & 125\% (13) & 117\% (12)\\
  \rowcolor[HTML]{EFEFEF} 
  TPCG1 & 43\% & 43\% & 44\% & 44\% \\
  TPCG2 & 61\% & 62\% & 63\% & 63\% \\
  \hline
\end{tabular}
\caption{Average time for the computation of the polarization energy and the associated forces for different methods, using the PCG converged at $10^{-5}$~D as reference, for a RESPA(2fs) timestep. In parenthesis, mean number of iterations needed.}
\end{table}


 We observe that both the TPCG methods are significantly faster compared to standard production settings (10-5D). 
Compared to more strict settings using a convergence criterion of $10^{-8}$~D for the PCG solver, which guarantees energy conservation during the MD simulation, differences are even more striking because the computational cost of the TPCG1 and TPCG2 methods are found to be respectively more than three times faster and more than twice faster respectively.

This means that using these methods with the implementation described in this paper enables not only to guarantee energy conservation but also to save a considerable amount of time during the computation of the polarization energy and the associated forces.

Concerning the use of ASPC, a striking result at a timestep of 2~fs is the smaller reduction of iterations necessary to reach convergence compared to the reduction observed at 1~fs \cite{Lipparini2014ScalableComputations} where a 50\% gain was observed for a $10^{-5}$D threshold.
In other words, ASPC guess is less efficient when using a bigger timestep.
Following intuition, the shorter the timestep, the more efficient the ASPC.
Moreover, in line with our previous study\cite{Lipparini2014ScalableComputations} , we also observed that the proportional gain in that regard is even smaller for tighter dipole convergence criterion (such as $10^{-8}$~D), making very long simulations a daunting challenge. 

Another remark concerns the use of even larger timesteps with the RESPA integrator.
It has been indeed shown that one can use a 3~fs timestep for the non-bonded forces, provided that masses of the hydrogen atoms of the system are appropriately redistributed among heavy atom carriers.\cite{hopkins2015long}
But such large timesteps limit the use of predictor such as the ASPC and no gain in the number of iteration can be obtained with these methods.
On the contrary, the computational cost of the T(P)CG family of methods does not suffer from such a change as no history is taken into account. The computational cost at 3 fs would remain the same that in the 2~fs context, offering an automatic 1.5 acceleration for the same trajectory length at no cost, increasing the global speedup offered by the use of T(P)CG.

\section{Conclusion}
As we have seen, one can reformulate the analytical expressions for the gradients of the Truncated Conjugate Gradient using a clear strategy. We detailed for interested developers the various steps required for the implementation of the complete TCG method including fast forces computations. 

This strategy allows the implementation of these gradients to be fast enough for the computational cost of an evaluation of the polarization energy and the associated forces to be greatly reduced compared to standard production settings using iterative methods. 
The TPCG2 method is more than 1.6 times faster than the PCG solver with a $10^{-5}$~D convergence criterion and the direct field as a guess using a RESPA integrator with a 2~fs time step (1.4 when ASPC is used).
Moreover, it is more than 2 times faster than a PCG with a convergence criterion of $10^{-8}$~D and the same predictor guess, such settings being mandatory to guarantee energy conservation with standard PCG for long simulations.
As the number of operations in the TCG method is fixed and does not rely on history (i.e. no previous dipole guess nor predictor guess), it can be applied with larger time-steps for the same fixed computational cost.

The TCG approach provides an accurate reproduction of energy surfaces \cite{Aviat2016TruncatedSimulations} at a reduced computational cost, providing analytical forces that avoid by construction the drift issues without relying on complex parametrization, nor adding extra degrees of freedom limiting the settings than one can use to integrate MD trajectories.
That is why it should be a method of choice for long timescale and stable simulations using polarizable force fields.
Since all TCG's analytical formulas involve the expressions of electric fields as well as matrix-vector products, these latter are easily and directly transposable in different boundary conditions. In particular, the extension to Smooth Particle Mesh Ewald is straightforward.
For the same reasons, the parallel implementation of these methods within the context of spatial decomposition follow any PCG one and will be described in a future paper dedicated to the massively parallel Tinker-HP package. 
In that context, capabilities of the AMOEBA force field using a TCG/SPME coupling will be tested by comparing various properties obtained with these methods.
\begin{acknowledgement}
This work was supported in part by French state funds managed by CalSimLab and the ANR within the Investissements d Avenir program under reference ANR-11-IDEX-0004-02. Jean-Philip Piquemal and Louis Lagardère are grateful for support by the Direction Générale de l Armement (DGA) Maitrise NRBC of the French Ministry of Defense.
\end{acknowledgement}
 
 \section{Annex\label{sec:Annex}}
 	We introduce the following notations to express the analytical formulas of the induced dipoles, as well as their derivatives. Each term can be expressed using the starting vectors ($\br_0$ and $\bmu_0$) and the polarization matrix $\bT$. 
    
    Vectors : 
    \begin{multicols}{2}
        \begin{itemize}
            \item $\br_0 = \bE - \bT \bmu_0 $
            \item $\bp_0 = \br_0 $ 
            \item $\bP_1 = \bT \br_0 $
            \item $\bP_2 = t_2 \bP_1 - t_4 \bT^2 \br_0$
            \item $\bP_3 = (1+\beta_2 t_2)\bT\br_0 - (t_4 + \beta_2 t_4) \bT\bP_1 - \gamma_1\bT \bP_2$
        \end{itemize}
    \end{multicols}
    
    Scalars :
    \begin{multicols}{3}
        \begin{itemize}
            \item $n_0 = \br_0^T \br_0$
            \item $t_1 = \br_0^T \bP_1$
            \item $\begin{aligned}[t]
                    t_2 = \frac{n_0 ||\bP_1||^2}{t_1^2}
                \end{aligned}$
            \item $t_3 = t_1 \bP_1^T \bP_2$
            \item $\begin{aligned}[t]
                    t_4 = \frac{n_0}{t_1}
                \end{aligned}$
            \item $t_5 = \bP_1^T \bP_2$
            \item $t_8 = t_5 = t_2 ||\bP_1||^2 - t_4 t_9$
            \item $t_9 = \br_0^T \bT^3 \br_0 $
            \item $t_{10} = t_1^2  - n_0 ||\bP_1||^2 $
            \item $\begin{aligned}[t]
                        \gamma_1 = \frac{ t_1^2 - n_0 ||\bP_1||^2 }{t_3}
                    \end{aligned}$
            \item $sp_0 = \br_0^T \bE$
            \item $sp_1 = \bP_1^T \bE = \bE^T \bT \br_0$
            \item $b_1 = sp_0 - \gamma_1 sp_1 $
            \item $b_2 = sp_0 t_2 - t_4 sp_1$ 
            \item $spp_1 = \langle \balpha \bE, \bE \rangle $
            \item $spp_2 = \langle \balpha \bT \br_0, \bE \rangle$
        \end{itemize}
    \end{multicols}
    	\begin{itemize}
    		\item $\begin{aligned}[t]
                    \beta_2 = \frac{n_0 + t_4^2 ||\bP_1||^2 + \gamma_1^2 ||\bP_2||^2 - 2 t_1 t_4  - 2 \gamma_1 t_4 ||\bP_1||^2 + 2 \gamma_1 t_4 t_5}{(t_2 - 1) n_0}
                \end{aligned}$
            \item $\begin{aligned}[t]
                \gamma_2 = \frac{n_0 + t_4^2 ||\bP_1||^2 + \gamma_1^2 ||\bP_2||^2 - 2 t_1 t_4  - 2 \gamma_1 t_4 ||\bP_1||^2 + 2 \gamma_1 t_4 t_5}{(1 + \beta_2 t_2) \br_0^T \bP_3 - (t_4 + \beta_2 t_4) \bP_1^T \bP_3 + \gamma_1 \bP_2^T \bP_3}
            \end{aligned}$
        \end{itemize}

 \subsection{Peek-step formulas}
    \begin{eqnarray}
        \bmu_{\text{peek, TCG1}} &=& \omega \balpha \br_0 - \omega  t_4 \balpha \bP_1 \\
        \bmu_{\text{peek, TCG2}} &=& \omega \balpha \br_0 - \omega t_4 \balpha \bP_1 - \omega \balpha \gamma_1 t_2 \bP_1 - \omega \balpha \gamma_1 t_4 \bT^2 \br_0 
    \end{eqnarray}

 \subsection{Coefficients for the analytical expressions}
    The superscript number, between parenthesis, indicates the truncation number (1 or 2).
    A $p$ indicates that the coefficient corresponds to the peek-step derivative, and needs to be added to the energy derivative coefficient itself. 
    
    Derivation of $E_{\text{pol, TCG1}}$ :
    \begin{multicols}{2}
    	\begin{itemize}
        	\item $a_{1,0}^{(1)} = t_4 $
            \item $a_{1,1}^{(1)} =  \frac{2 sp_0}{t_1} + t_4 $
            \item $a_{1,2}^{(1)} =  -\frac{ 2 sp_0 n_0}{t_1^2} $
            \item $a_{2,1}^{(1)} =  - \frac{sp_0 n_0 }{t_1^2} $
        \end{itemize}
    \end{multicols}
    Peek-step for  TCG1 :
    \begin{multicols}{2}
    	\begin{itemize}
          \item $ a_{1,\alpha0}^{(1,p)} = \omega  $ 
          \item $ a_{1,1\alpha}^{(1,p)} = -t_4 \omega  $ 
          \item $ a_{1,1}^{(1,p)} = - \frac{2spp_1 \omega}{t_1}  $ 
          \item $ a_{1,2}^{(1,p)} = \frac{2n_0 spp_1 \omega}{t_1^2}  $ 
          \item $ a_{2,\alpha0}^{(1,p)} = -t_4 \omega  $ 
          \item $ a_{2,1}^{(1,p)} = \frac{n_0 spp_1 \omega}{t_1^2} $
         \end{itemize}
     \end{multicols}
     TCG2 :
 	\begin{multicols}{1}
    	\begin{itemize}
        	\item $a_{1,0}^{(2)} =  t_4 + \gamma_1 t_2 $
            \item $a_{1,-1}^{(2)} = -\gamma_1 t_4 $
            \item $a_{1,1}^{(2)} = \frac{2 b_1}{t_1} - \frac{2 np_1 b_2}{t_3} - 2 \frac{np_1^2 t_{10} b_2}{t_3^2 t_1} + 2 \frac{t_9 t_{10} b_2}{t_3^2} + 2 \frac{np_1 sp_0 \gamma_1}{t_1^2} $
            \item $a_{1,2}^{(2)} = -2 \frac{n_0 b_1}{t_1^2} + 4 \frac{t_1 b_2}{t_3} - 2 \frac{n_0 t_9 t_{10} b_2}{t_1 t_3^2} + 4 \frac{t_2 np_1 t_{10} b_2}{t_3^2} - 2 \frac{t_8 t_{10} b_2}{t_3^2} - 4 \frac{4 n_0 np_1 sp_0 \gamma_1}{t_1^3} $
            \item $a_{1,3}^{(2)} = -4 \frac{t_1 t_2 t_{10} b_2}{t_3^2} - 2 \frac{n_0 b_2}{t_3} + 2 \frac{n_0 sp_0 \gamma_1}{t_1^2}  $
            \item $a_{1,4}^{(2)} = 2 \frac{t_1 t_4 t_{10} b_2}{t_3^2} $
            \item $a_{2,0}^{(2)} = -\gamma_1 t_4 $
            \item $a_{2,1}^{(2)} = - \frac{n_0 b_1}{t_1^2} + 2 \frac{t_1 b_2}{t_3} - \frac{n_0 t_9 t_{10} b_2}{t_1 t_3^2} + 2 \frac{t_2 np_1 t_{10} b_2}{t_3^2} - \frac{t_8 t_{10} b_2}{t_3^2} - 2 \frac{n_0 np_1 sp_0 \gamma_1}{t_1^3} $
            \item $a_{2,2}^{(2)} = -\frac{n_0 b_2}{t_3} - 2 \frac{t_1 t_2 t_{10} b_2}{t_3^2} + \frac{n_0 sp_0 \gamma_1}{t_1^2} $
            \item $a_{2,3}^{(2)} = \frac{t_1 t_4 t_{10} b_2}{t_3^2} $
            \item $a_{3,1}^{(2)} = - \frac{n_0 b_2}{t_3} - 2 \frac{t_1 t_2 t_{10} b_2}{t_3^2} + \frac{n_0 \gamma_1 sp_0}{t_1^2} $
            \item $a_{4,1}^{(2)} = \frac{t_1 t_4 t_{10} b_2}{t_3^2} $
            \item $a_{3,2}^{(2)} = \frac{t_1 t_4 t_{10} b_2}{t_3^2} $
        \end{itemize}
    \end{multicols}

    Peek-step for TCG2 :
    \begin{multicols}{2}
    	\begin{itemize}
        	\item $a_{1,0\alpha}^{(2,p)} = \omega $
        	\item $a_{1,1\alpha}^{(2,p)} = - \omega (t_2 \gamma_1 + t_4)  $
            \item $a_{1,2\alpha}^{(2,p)} = -\omega t_4 \gamma_1 $
        \end{itemize}
    \end{multicols}
    	\begin{itemize}
            \item $a_{1,1}^{(2,p)} = -\frac{2 np_1 }{t_1^2} \omega \gamma_1 spp_1 + \left( \omega t_2 spp_1 + \omega t_4 spp_2 \right) \left( \frac{2 np_1}{t_3} + \frac{2 np_1^2 t_{10}}{t_1 t_3^2} - \frac{2 t_9 t_{10}}{t_3^2} \right) - \frac{2  }{t_1} (\omega \gamma_1 spp_2 + \omega spp_1) $
            \item $a_{1,2}^{(2,p)} =\frac{4 n_0 np_1 }{t_1^3} \omega \gamma_1 spp_1 + (\omega t_2 spp_1 + \omega t_4 spp_2) \left( -\frac{4 t_1}{t_3} + \frac{2 n_0 t_9 t_{10}}{t_1 t_3^2} - \frac{4 np_1 t_2 t_{10}}{t_3^2} + \frac{2 t_8 t_{10}}{t_3^2}\right) + \frac{2 n_0 }{t_1^2} (\omega \gamma_1 spp_2 + \omega spp_1) $
            \item $a_{1,3}^{(2,p)} = - \frac{2 n_0 }{t_1^2} \gamma_1 \omega spp_1 + (\omega t_2 spp_1 + \omega t_4 spp_2) \left( \frac{4 t_1 t_2 t_{10}}{t_3^2} + \frac{2 n_0}{t_3} \right) $
        \end{itemize}
   \begin{multicols}{2}
   		\begin{itemize}
            \item $a_{1,4}^{(2,p)} = - (\omega t_2 spp_1 + \omega t_4 spp_2) \frac{2 t_1 t_4 t_{10}}{t_3^2} $
            \item $a_{2,\alpha0}^{(2,p)} = -\omega (\gamma_1 t_2 + t_4) $
            \item $a_{2, 1\alpha}^{(2,p)} =- \omega t_4 \gamma_1 $
        \end{itemize}
    \end{multicols}
    	\begin{itemize}
            \item $a_{2,1}^{(2,p)} =\frac{2 n_0 np_1 }{t_1^3}  \omega \gamma_1 spp_1 + (\omega t_2 spp_1 + \omega t_4 spp_2) \left( -\frac{2 t_1}{t_3} + \frac{n_0 t_9 t_{10}}{t_1 t_3^2} - \frac{2 np_1 t_2 t_{10}}{t_3^2} + \frac{t_8 t_{10}}{t_3^2} \right) + \frac{n_0 }{t_1^2} (\omega \gamma_1 spp_2 + \omega spp_1) $
            \item $a_{2,2}^{(2,p)} = - \frac{n_0 }{t_1^2} \omega \gamma_1 spp_1 + (\omega t_2 spp_1 + \omega t_4 spp_2) \left( \frac{n_0 }{t_3} + \frac{2 t_1 t_2 t_{10}}{t_3^2} \right) $
        \end{itemize}
    \begin{multicols}{2}
   		\begin{itemize}
            \item $a_{2,3}^{(2,p)} = - (\omega t_2 spp_1 + \omega t_4 spp_2) \frac{t_1 t_4 t_{10}}{t_3^2} $
            \item $a_{3,\alpha0}^{(2,p)} = - \omega  \gamma_1 t_4 $
        \end{itemize}
     \end{multicols}
     	\begin{itemize}
            \item $a_{3,1}^{(2,p)} = - \frac{n_0 }{t_1^2} \omega \gamma_1  spp_1 + (\omega t_2 spp_1 + \omega t_4 spp_2 ) \left( \frac{n_0}{t_3} + \frac{2 t_1 t_2 t_{10}}{t_3^2} \right) $
        \end{itemize}
      \begin{multicols}{2}
      	  \begin{itemize}
            \item $a_{3,2}^{(2,p)} = -  (\omega t_2 spp_1 + \omega t_4 spp_2) \frac{t_1 t_4 t_{10}}{t_3^2} $
            \item $a_{4,1}^{(2,p)} = - (\omega t_2 spp_1 + \omega t_4 spp_2) \frac{t_1 t_4 t_{10}}{t_3^2} $
        \end{itemize}
    \end{multicols}



\bibliography{TCGfastgrad.bib}

\end{document}